\documentclass[%
 reprint,
 amsmath,
 amssymb,
 aps,
 pra,
 showpacs
]{revtex4-1}

\usepackage{graphicx}
\usepackage{dcolumn}
\usepackage{bm}
\usepackage[driverfallback=dvipdfmx]{hyperref}
\usepackage{braket}

\newcommand{\w}{\omega}
\newcommand{\ka}{\kappa}
\newcommand{\s}{\sigma}
\newcommand{\f}[2]{\frac{#1}{#2}}
\newcommand{\M}{\mathcal{M}}
\newcommand{\D}{\mathrm{d}}

\begin{document}

\title{Non-Markovian master equation for interacting qubits coupled to a bosonic bath: analytic form and asymptotic approximation}

\author{Tai-Yin Chiu}
\email{b99202046@ntu.edu.tw}
\affiliation{Graduate Institute of Electronics Engineering, National Taiwan University, Taipei, Taiwan}

\begin{abstract}
Non-Markovian dynamics of two interacting two-level qubits coupled to a bosonic bath was previously studied using the quantum-state-diffusion (QSD) equation, where a stochastic state is used to describe the system. In this study, we provide another perspective on this system by deriving the analytic form of the master equation, which describes the system with a reduced density matrix. Then, we validate the master equation by examining entanglement generation and state purity. In addition, with the master equation, we observe the effects from first-order noise and notice that a good asymptotic approximation to the master equation can be made by neglecting the first-order noise.
\end{abstract}


\maketitle

\section{Introduction}
\label{sec:intro}
In the real world, a quantum system will always interact with the external environment and cannot be treated as an isolated system, leading to the introduction of open quantum systems \cite{Breuer2007}. Recently, open quantum systems have attracted significant attention, and phenomena such as quantum decoherence \cite{Joos2003,Schlosshauer2005,Helm2009} and quantum dissipation \cite{Diehl2008,Krauter2011,Weiss2012} have always been a key focus of several researchers in this field. Moreover, open quantum system concepts have been extensively applied in quantum optics \cite{Carmichael2014}, quantum measurement and metrology \cite{Schlosshauer2005,Barchielli1983,Goldstein2011,Alipour2014}, quantum control\cite{Liu2011,Schmidt2011}, and quantum information science including quantum computation \cite{Verstraete2009} and quantum simulation\cite{Barreiro2011,Houck2012}.

Typically, an open quantum system is described by a reduced density matrix, and its dynamics is mathematically governed by a master equation. In a Markov approximation, the time evolution of a system is given by the well-known Lindblad master equation \cite{Lindblad1976}. However, the Markov limit is not always a good approximation since open systems in Markov limit usually lose information to the surroundings and exhibit irreversible dissipation and decoherence, which are bottlenecks to address in quantum information science. On the contrary, non-Markovian process-based systems \cite{Breuer2007,Cederbaum2005,Apollaro2011}, where back reaction of environments is considered, regain the information previously lost to the surroundings. This makes the non-Markovian processes more physical and appropriate than the Markovian process under certain circumstances. One useful approach to cope with non-Markovian situations is called the non-Markovian stochastic Schr\"{o}dinger equation or the quantum-state-diffusion (QSD) equation \cite{Diosi1998,Diosi1998_2}; it describes an open system by a stochastic pure state rather than a density matrix. 

One of the successful cases that was dealt with using QSD is a system comprising two interacting qubits coupled to a bosonic environment \cite{Zhao2011}, which was studied by Zhao \emph{et~al.} They derived the exact time-local QSD equation for such a system and discovered the entanglement generation caused by the environmental memory. However, describing an open system with a reduced density matrix instead of a stochastic pure state would sometime be better since a stochastic state may carry an uncertainty represented by a complex stochastic Gaussian process. In contrast, the uncertainty in a reduced density matrix is ruled out by an ensemble average. In addition, in quantum control problems such as those discussed in \cite{Tai2014}, knowledge of the analytic form of the master equation is quite important. Furthermore, since control problems usually require a large amount of calculation, making a good approximation to reduce the computational complexity is a necessity. 

Based on the formal form of the master equation for multi-qubit dissipative systems \cite{Chen2014}, we have derived the analytic form of the master equation for this two-qubit system, where the noise is modeled by the Ornstein\textendash Uhlenbeck process. In particular, we write down an exact master equation composed of several functions whose time derivatives are analytically given. Moreover, we validate our analytic master equation by examining the entanglement generation and the purity of state, comparing them to the results in \cite{Zhao2011}. In addition, we discuss the effects of first-order noise and show that from an asymptotic perspective, the exact master equation can be reduced to an approximate form, wherein the terms related to the first-order noise are eliminated. 

This study is organized as follows. In Sec.~\ref{sec:math_des}, we introduce the model and the stochastic Schr\"{o}dinger equation for the system comprising two interacting atoms coupled to a bosonic environment and derive the analytic form of the non-Markovian master equation. In Sec.~\ref{sec:sim_discuss}, we use the non-Markovian master equation to examine the entanglement generation and state purity phenomena due to environmental memory and then validate the master equation. Next, we compare the solutions of the exact and approximate master equations in the last part of this section. Finally, Sec.~\ref{sec:conclusion} presents a conclusion of this study.

\section{Mathematical Description}
\label{sec:math_des}
\subsection{Model and stochastic Schr\"{o}dinger equation}
For a system comprising two interacting two-level atoms coupled to a bosonic bath \cite{Zhao2011}, the total Hamiltonian can be divided into three parts (set $\hbar =1$):
\begin{equation} 
\label{eq:hamiltonian}
	\begin{aligned}
	H_{sys} = &~\w_A \s_z^A + \w_B \s_z^B + J_{xy}(\s_+^A \s_-^B + \s_-^A \s_+^B)\\ &+ J_z \s_z^A \s_z^B, \\
	H_{bath} =& \sum_i \w_i a_i^\dag a_i, \\
	H_{int} =& \sum_i (g_i a_i^\dag L + g_i^\ast a_i L^\dag),
	\end{aligned}
\end{equation}
where $w_A$ and $w_B$ are the transition frequencies of two interacting atoms, $\s_\pm = (\s_x \pm i\s_y)/2$ is the creation/annihilation operator for an atom, $a_i (a_i^\dag)$ is the annihilation (creation) operator of the $i^{\text{th}}$ mode of the bosonic bath, $g_i$ is the coupling constant between the system and the $i^{\text{th}}$ mode of the bath, and $L = \ka_A \s_-^A + \ka_B \s_-^B$ is a system operator. Note that the interaction between two atoms is modeled on the Heisenberg XXZ model, wherein the coupling constants $J_x = J_y = J_{xy}$ are not necessarily equal to $J_z$.

The non-Markovian stochastic Schr\"{o}dinger equation for two interacting atoms coupled to a common bath is given as follows:
\begin{equation} 
\label{eq:SSE}
\f{\partial}{\partial t} \psi_t = -i H_{sys} \psi_t + Lz_t^\ast \psi - L^\dag\int_0^t \D s \alpha(t,s) \f{\delta \psi_t}{\delta z_s^\ast},
\end{equation}
where $\psi_t = \psi_t(z^\ast)$ is a stochastic wave function of the system, and $z_t \equiv i \sum_j g_j z_j e^{-i\w_j t}$ is a complex Gaussian process satisfying $\M[z_t] = 0$, $\M[z_t z_s] = 0$, and $\M[z_t z_s^\ast] = \alpha(t,s)$, which is the correlation function of the bath and determines the environmental memory time. Here, the symbol, $\M[\bullet] = \int \cdots \int \f{\D^2 z_j}{\pi} e^{-|z_j|^2} \cdots \int \f{\D^2 z_1}{\pi} e^{-|z_1|^2}(\bullet)$, represents the ensemble average operation.

In general, the functional derivative in the integral of Eq.~(\ref{eq:SSE}) can be replaced with an operator $O(t,s,z^\ast)$, such that
\begin{equation}
\label{eq:diff2op}
\f{\delta \psi_t(z^\ast)}{\delta z_s^\ast} = O(t,s,z^\ast)\psi_t(z^\ast).
\end{equation}
For this two-interacting-qubit model, the $O$ operator can be written as follows:
\begin{equation}
\label{eq:O_op_form}
O(t,s,z^\ast) = O_0(t,s) + \int_0^t \D s_1 z_{s_1}^\ast O_1(t,s,s_1),
\end{equation}
where $O_0(t,s)$ and $O_1(t,s,s_1)$ are related to the zeroth- and first-order noise components, which may be formulated as follows: 
\begin{equation}
\label{eq:O_op}
\begin{aligned}
O_0(t,s) =&~ f_1(t,s) \s_-^A + f_2(t,s) \s_-^B + f_3(t,s) \s_z^A \s_-^B \\
&+ f_4(t,s) \s_-^A \s_z^B, \\
O_1(t,s,s_1) =&~ i f_5(t,s,s_1)(2\s_-^A \s_-^B),
\end{aligned}
\end{equation}
in which the time derivatives and the initial conditions of the $f_j$ terms are listed in Appendix \ref{app:time_deri}.

\subsection{Exact non-Markovian master equation}
In general, the formal non-Markovian master equation can be derived from Eq.~(\ref{eq:SSE}), and it reads as follows:
\begin{equation}
\label{eq:formal_ME}
\begin{aligned}
\f{\partial \rho_t}{\partial t} =& -i[H_{sys},\rho_t] + \big[L,\M[P_t\bar{O}^\dag(t,z^\ast)]\big] \\ &+ \big[M[\bar{O}(t,z^\ast)P_t],L^\dag\big],
\end{aligned}
\end{equation}
where $\bar{O}(t,z^\ast) = \int_0^t \D s\alpha(t,s)O(t,s,z^\ast)$, $P_t = \Ket{\psi_t(z^\ast)}\Bra{\psi_t(z)}$, and $\rho_t = \M[P_t]$, which is the reduced density matrix used to describe the dynamics of the system.
For a system that has the first order of noise as the highest order, $\M[P_t \bar{O}^\dag]$ can be written as follows\cite{Chen2014}:
\begin{widetext}
\begin{equation}
\label{eq:pt_obar_dag}
\begin{aligned}
\M[P_t \bar{O}^\dag] =&~ \rho_t \bar{O}_0^\dag(t) + \int_0^t \D s_1 \int_0^t \D s_2 \alpha(s_1,s_2) O_0(t,s_2) \rho_t \bar{O}_1^\dag(t,s_1) 
\\ &+ \int_0^t \D s_1 \int_0^t \D s_2 \int_0^t \D s_3 \int_0^t \D s_4 \alpha(s_1,s_2)\alpha(s_3,s_4)^\ast O_1(t,s_2,s_3) \rho_t O_0^\dag(t,s_4) \bar{O}_1^\dag(t,s_1),
\end{aligned}
\end{equation}
\end{widetext}
where $\bar{O}_0(t) = \int_0^t \D s \alpha(t,s) O_0(t,s)$ and $\bar{O}_1(t,s_1) = \int_0^t \D s \alpha(t,s) O_1(t,s,s_1)$. In our two-qubit system, the term involving the quadruple integral vanishes since $O_0^\dag\bar{O}_1 = 0$ due to $\s_+\s_+ = 0$.

To make good use of Eq.~(\ref{eq:formal_ME}), we have to convert Eq.~({\ref{eq:pt_obar_dag}}) to a more explicit form, from which $\M[P_t\bar{O}^\dag]$ can be easily computed. For this, we first define
\begin{equation}
\bar{f}_j(t) \equiv \int_0^t \D s~\alpha(t,s)f_j(t,s),~ j = 1 \sim 4,
\end{equation}
and
\begin{equation}
\begin{aligned}
F_j(t) \equiv \int_0^t \D s_1 \int_0^t \D s_2~\alpha(s_1,s_2)f_j(t,s_2)&\bar{f}_5^\ast(t,s_1),\\& j = 1 \sim 4. 
\end{aligned}
\end{equation}
When $O_0$ and $O_1$ given in Eq.~(\ref{eq:O_op}) are substituted into Eq.~(\ref{eq:pt_obar_dag}), the equation gets converted to an explicit form as shown below:
\begin{equation}
\label{eq:pt_obar_fun}
\begin{aligned}
\M&[P_t\bar{O}^\dag] = \rho_t(\bar{f}^\ast_1 \s_+^A + \bar{f}^\ast_2 \s_+^B + \bar{f}^\ast_3 \s_z^A \s_+^B + \bar{f}^\ast_4 \s_+^A s_z^B) \\
&-2i[F_1 \s_-^A + F_2 \s_-^B + F_3 \s_z^A \s_-^B + F_4 \s_-^A \s_z^B)\rho_t \s_+^A \s_+^B.
\end{aligned}
\end{equation}
Here, we consider the Ornstein\textendash Uhlenbeck process with the correlation function $\alpha(t,s) = \f{\gamma}{2}e^{-\gamma|t-s|}$, which corresponds to a Lorentzian power spectrum, to model the noise. In this process, with the help of the three other auxiliary functions
\begin{equation}
F_5(t) = \int_0^t \D s_1 \int_0^t \D s_2~\alpha(s_1,s_2)\bar{f}_5(t,s_2)\bar{f}_5^\ast(t,s_1),
\end{equation}
\begin{equation}
\bar{f}_5(t,s_1) = \int_0^t \D s~\alpha(t,s)f_5(t,s,s_1), 
\end{equation}
and
\begin{equation}
\tilde{f}_5(t) = \int_0^t \D s~\alpha(t,s) \bar{f}_5(t,s),
\end{equation}
the explicit form of $\M[P_t\bar{O}^\dag]$ in Eq.~(\ref{eq:pt_obar_fun}) becomes very useful, since $\big\{\f{\D \bar{f}_1}{\D t}, \f{\D \bar{f}_2}{\D t}, \f{\D \bar{f}_3}{\D t}, \f{\D \bar{f}_4}{\D t}, \f{\D \tilde{f}_5}{\D t}\big\}$ and $\big\{\f{\D F_1}{\D t}, \f{\D F_2}{\D t}, \f{\D F_3}{\D t}, \f{\D F_4}{\D t}, \f{\D F_5}{\D t}\big\}$ form two sets of coupled differential equations (see Appendix \ref{app:time_deri}), implying that the analytic $\M[P_t \bar{O}^\dag]$ can be easily calculated by numerically solving these differential equations. With the derivation of $\M[P_t\bar{O}^\dag]$, we can compute $\rho_t$ from Eq.~({\ref{eq:formal_ME}), and collectively, this completes the construction of the analytic form of the exact non-Markovian master equation for this two-interacting-atom system.

\subsection{Markovian regime and noise effects}
If the correlation function $\alpha(t,s)$ in Eq.~(\ref{eq:SSE}) is given by $\gamma\delta(t-s)$, meaning that the quantum state at time $t$ is independent of the state at time $s$ ($s < t$), then there is no memory effect in the surrounding, and Eq.~({\ref{eq:SSE}) becomes
\begin{equation} 
\label{eq:SSE_markovian}
\f{\partial}{\partial t} \psi_t = -i H_{sys} \psi_t + Lz_t^\ast \psi_t - \f{\gamma}{2}L^\dag L \psi_t,
\end{equation}
which is the stochastic Schr\"{o}dinger equation in the Markovian limit. Note that in the Ornstein\textendash Uhlenbeck process, when $\gamma \to \infty$, the correlation function will behave like a Dirac delta function, and hence, the environment becomes memoryless.

In fact, when the system is close to the Markovian regime, the operator $O_1(t,s,s_1)$ associated with the first-order noise in Eq.~({\ref{eq:O_op_form}) becomes less important. To find the noise effect from $O_1$, we may approximate $O(t,s,z^\ast)$ with $O_0(t,s)$ or equivalently let $f_5(t,s,s_1) = 0$ and then compare the results from the exact and approximate master equations. In the approximate master equation, $M[P_t\bar{O}^\dag]$ is reduced to
\begin{equation}
\M[P_t\bar{O}^\dag] = \rho_t(\bar{f}^\ast_1 \s_+^A + \bar{f}^\ast_2 \s_+^B + \bar{f}^\ast_3 \s_z^A \s_+^B + \bar{f}^\ast_4 \s_+^A s_z^B),
\end{equation}
where the time derivatives of the $\bar{f}_j$ terms form a set of coupled differential equations obtained by eliminating $\bar{f}_5$ and $\tilde{f}_5$ from Eq.~(\ref{eq:dfbar_dt}).

\section{Numerical simulations and discussions}
\label{sec:sim_discuss}
In this section, we validate the newly derived master equation and compare the time evolutions of the density matrix $\rho_t$ governed by the exact and approximate $O$ operators. Here, the states and operators are represented in the basis $\{\Ket{11}, \Ket{10}, \Ket{01}, \Ket{00} \}$, where $\s_z\Ket{1} = \Ket{1}$ and $\s_z\Ket{0} = -\Ket{0}$.

\subsection{Entanglement generation and state purity}

To validate the exact master equation, we examine whether it can reproduce the same entanglement generation and time evolution of state purity as given by the QSD equation in \cite{Zhao2011}. Here purity is defined as $\text{tr}[\rho^2(t)]$, which conveys whether a system is in a pure or a mixed state.

From \cite{Zhao2011}, we know that if the initial state is $\psi_0 = \ket{10}$ (or $\rho_0 = \ket{10}\bra{10}$), the two-qubit system evolves into a mixed and entangled steady state; while, if the initial state is $\psi_0 = \f{1}{\sqrt{2}}(\ket{11}+\ket{00}), \f{1}{\sqrt{2}}(\ket{10}+\ket{01})$, or $\ket{11}$, the system will first evolve to a mixed state because of the coupling interaction and finally decay to the pure state $\ket{00}$ resulting from the dissipative environment.

\begin{figure}[htb]
\centering
\includegraphics[width=7cm]{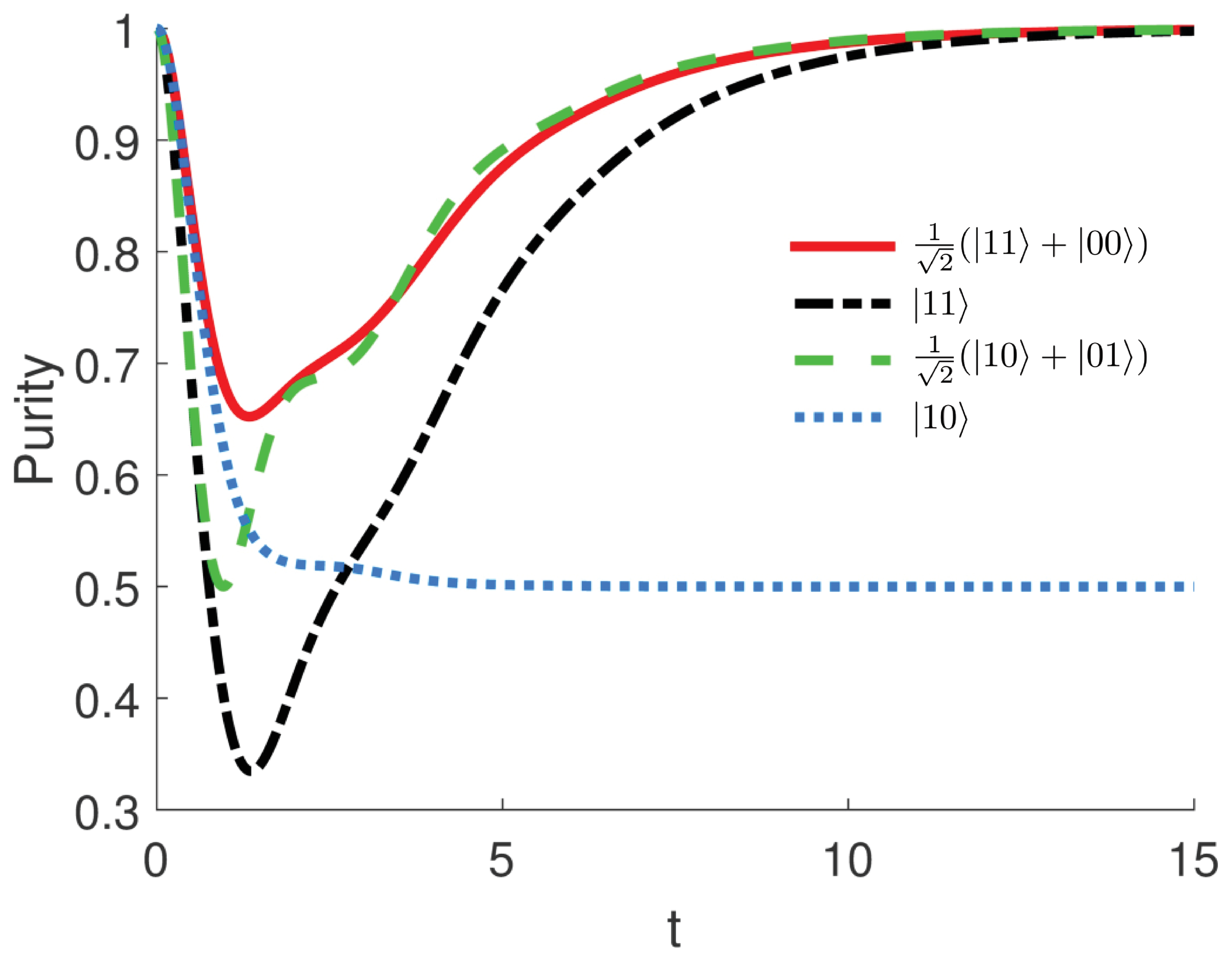}
\caption{(Color online) Time evolution of purity for four different initial states. Here $\gamma = 1$, $\w_A = \w_B = 0.5$, $\ka_A = \ka_B = 1$, $J_{xy} = 0.7$, and $J_z = 0.3$.}
\label{fig:purity_exact}
\end{figure}

In Fig.~\ref{fig:purity_exact}, we plot the time evolution of purity for those four initial states generated by the exact master equation. We can see  that unlike the other three initial states whose purities first become smaller than unity and then return to unity, the system remains in a mixed state with a purity value close to $0.5$ only when $\psi_0 = \ket{10}$. 

Moreover, from the time evolution of the density matrix with $\rho_0 = \ket{10}\bra{10}$, plotted in Fig.~\ref{fig:ent_purity}, we can clearly observe the system ending up in a state that is not only mixed but also entangled. Therefore, we can infer that the results from the exact master equation match well with the ones from the QSD equation described above, and this consistency confirms the validity of our master equation.

\begin{figure}[htb]
\centering
\includegraphics[width=8.6cm]{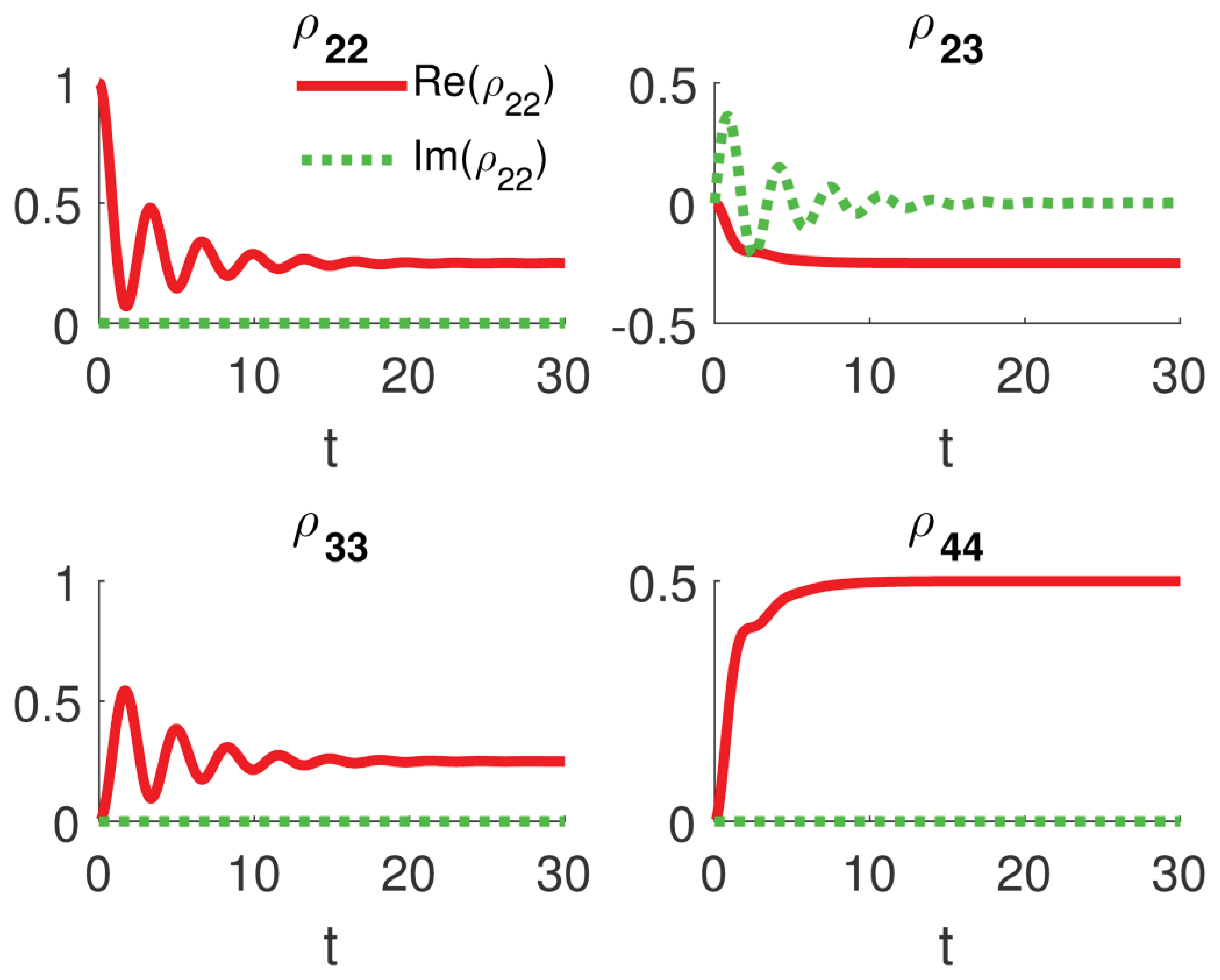}
\caption{(Color online) Dynamics of the density matrix with the initial state $\psi_0 = \ket{10}$. Parameters are chosen as $\gamma = 1$, $\w_A = \w_B = 0.5$, $\ka_A = \ka_B = 1$, $J_{xy} = 0.7$, and $J_z = 0.3$. The values of other elements are always zero except $\rho_{32} = \rho_{23}^\ast$.}
\label{fig:ent_purity}
\end{figure}

\begin{figure*}[bht]
\centering
\includegraphics[width=17.2cm]{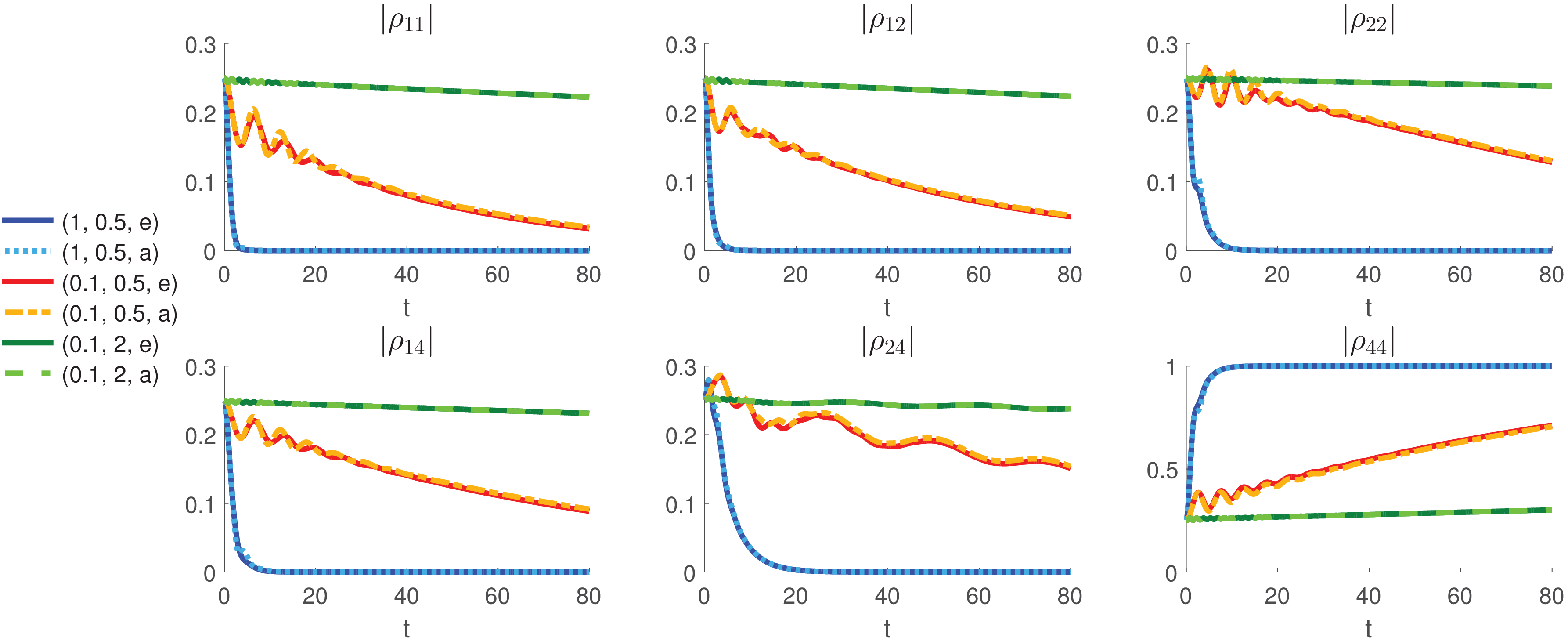}
\caption{(Color online) Dynamics of the density matrix with different combinations of $\gamma$ and $\w$. Here, the initial state is $\psi_0 = \f{1}{2}(\ket{11}+\ket{10}+\ket{01}+\ket{00})$. Legend ($\gamma,~\w,~$e/a) indicates values of $\gamma$ and $\w$ for a curve, and e/a is used to indicate whether a curve is generated using the exact or approximate master equation. Other parameters are chosen as $\ka = 1$, $J_{xy} = 0.7$, and $J_z = 0.3$. Note that the absolute value of each complex element is shown here.}
\label{fig:g_w_cmp}
\end{figure*}

\subsection{Exact master equation versus approximate master equation}

In the following, we assume that $\w_A = \w_B = \w$ and $\ka_A = \ka_B = \ka$ for simplicity. For such a condition, we can deduce from Eq.~({\ref{eq:dfbar_dt}) that $\bar{f}_1 = \bar{f}_2$ and $\bar{f}_3 = \bar{f}_4$ given the symmetry in their time derivatives and the same initial conditions, share $\bar{f}_1(0) = \bar{f}_2(0) = \bar{f}_3(0) = \bar{f}_4(0) = 0$. Furthermore, if we substitute $\psi_t = c_1(t) \Ket{11} + c_2(t) \Ket{10} + c_3(t) \Ket{01} + c_4(t) \Ket{00}$ into Eq.~(\ref{eq:SSE}), the dynamics of the $c_j$ values are found to be
\begin{equation}
\label{eq:dcj_dt}
\begin{aligned}
\f{\D c_1}{\D t} =& -i(2\w+J_z)c_1 - \ka(2\bar{f}_1+2\bar{f}_3)c_1, \\
\f{\D c_2}{\D t} =&~ \ka z_t^\ast c_1 - 2\ka \hat{f}_5 c_1 + i J_z c_2 - \ka(\bar{f}_1 - \bar{f}_3)c_2 -i J_{xy}c_3 \\ &- \ka(\bar{f}_1-\bar{f}_3)c_3, \\
\f{\D c_3}{\D t} =&~ \ka z_t^\ast c_1 - 2\ka \hat{f}_5 c_1 + i J_z c_3 - \ka(\bar{f}_1 - \bar{f}_3)c_2 -i J_{xy}c_2 \\ &- \ka(\bar{f}_1-\bar{f}_3)c_3, \\
\f{\D c_4}{\D t} =&~ \ka z_t^\ast (c_2 + c_3) - i(J_z - 2\w)c_4,
\end{aligned}
\end{equation}
where $\hat{f}_5(t,z^\ast) = i\int_0^t \D s_1 \bar{f}_5(t,s_1) z_{s_1}^\ast$. Again, we can infer the relationship $c_2(t) = c_3(t)$ when $c_2(0) = c_3(0)$ because of the symmetry between $\dot{c}_2$ and $\dot{c}_3$. For convenience, we shall assume $c_2(0) = c_3(0)$ without any loss of generality in the comparisons.

Under these assumptions, $\rho_{j2} = \M[c_j c_2^\ast] = \M[c_j c_3^\ast] = \rho_{j3}$ ($j = 1 \sim 4$), and thereby, together with the property $\rho_t = \rho_t^\dag$ of a density matrix, there are only six independent elements in $\rho_t$: $\rho_{11}$, $\rho_{12}$, $\rho_{14}$, $\rho_{22}$, $\rho_{24}$, and $\rho_{44}$. Next, we compare the differences of these six elements considering the situations with or without the first-order noise.

\begin{figure}[!htb]
\centering
\includegraphics[width=8.6cm]{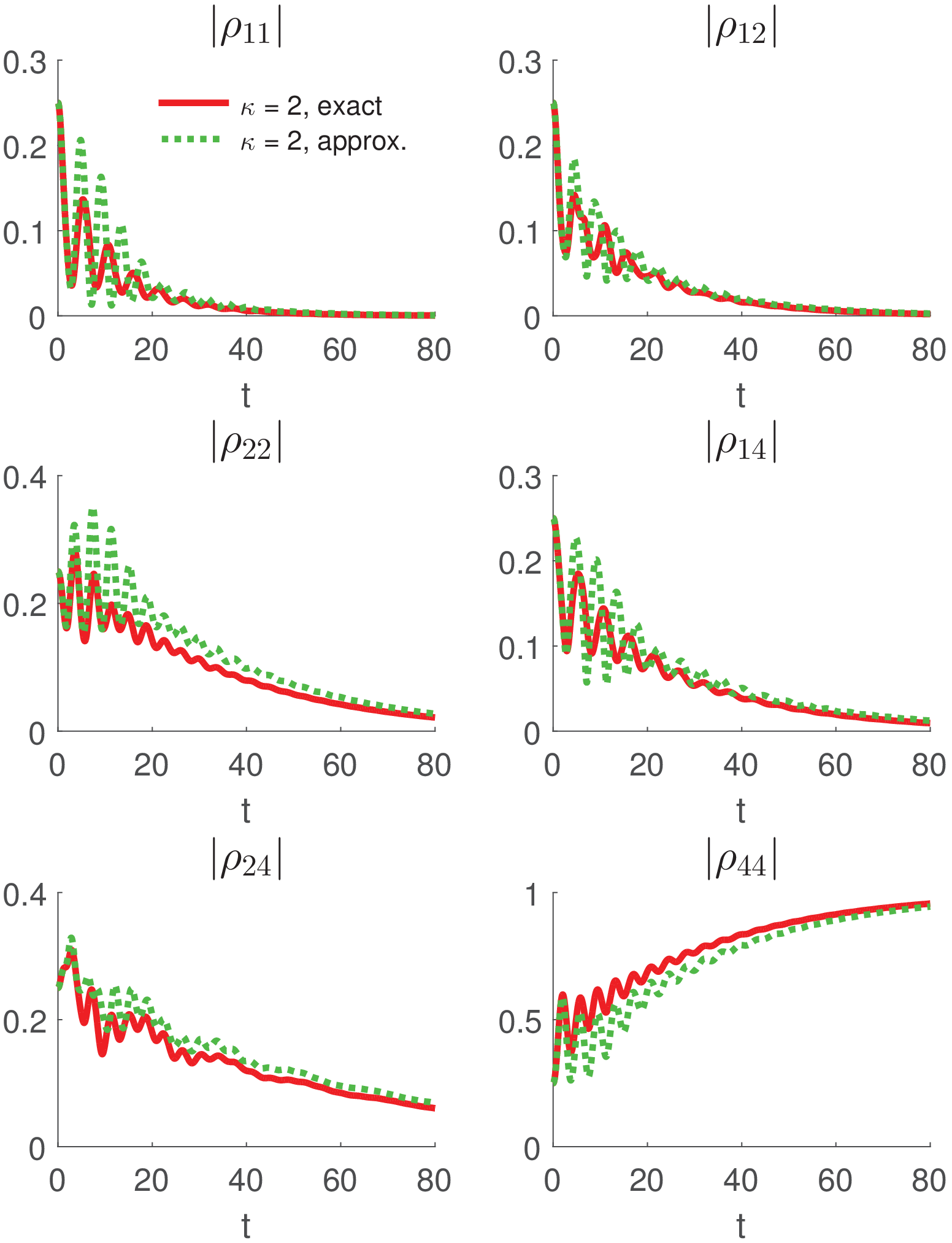}
\caption{(Color online) Dynamics of the density matrix with $\gamma = 0.1$, $\w = 0.5$, $\ka = 2$, $J_{xy} = 0.7$, and $J_z = 0.3$. The initial state is $\psi_0 = \f{1}{2}(\ket{11}+\ket{10}+\ket{01}+\ket{00})$.}
\label{fig:k_cmp}
\end{figure}

First, we consider the time evolution of a state, $\psi_t$, initialized as $\psi_0 = \f{1}{2}(\ket{11}+\ket{10}+\ket{01}+\ket{00})$ with parameters $\gamma = 0.1$, $\w = 0.5$, and $\ka = 1$. From Fig.~\ref{fig:g_w_cmp}, we can infer that the dynamics of $\psi_t$ governed by the exact and approximate master equations only show very small differences although the environment is way off from the Markovian regime because of the small value of $\gamma$. Next, if we increase the value of $\gamma$ to unity, it is as expected that the system reaches its final state faster due to limited back reaction from the environment in the Markovian regime. In this case, the results produced by both master equations match each other perfectly because the first-order noise is less important in the Markov limit. From Fig.~\ref{fig:g_w_cmp}, we can further deduce that the approximate master equation behaves well when $\w$ increases from $0.5$ to $2$, while the environment is still in a non-Markovian regime ($\gamma = 0.1$). In fact, variation of $J_{xy}$ and $J_z$ also does not initiate any clear difference between the results from both these master equations.

Next, in Fig.~\ref{fig:k_cmp}, we show the dynamics from the same initial state governed by the exact and approximate master equations when $\gamma = 0.1$, $\w = 0.5$, and $\ka = 2$. Compared with the red solid line and orange dashed line in Fig.~\ref{fig:g_w_cmp} with $\ka = 1$, the oscillations are more intense with a shorter response time for each element of the density matrix, which can be explained by the analytical solution to Eq.~(\ref{eq:dcj_dt}). For example, from the solution for $c_1(t)$ 
\begin{equation}
c_1(t) = c_1(0)e^{-i(2\w+J_z)t}e^{-2\ka\int_0^t (\bar{f}_1(\tau) + \bar{f}_3(\tau)) \D \tau},
\end{equation}
we can infer that $\ka$ has a direct influence on the oscillations and the response time. Incidentally, the same effect can also be observed in the solutions to $c_2$, $c_3$, and $c_4$.

Moreover, it is also observed that the results from the exact and approximate master equations do not match each other in the transient time. As for the cause, we first notice in Eq.~(\ref{eq:O_op}) that the operator associated with the first-order noise is $\s_-^A \s_-^B$, and then, apparently, the difference comes from the coefficient $c_1(t)$ of state $\ket{11}$, which is operated on by $\s_-^A \s_-^B$. To verify this argument, we consider another initial state $\psi_0 = \f{1}{\sqrt{3}}(\ket{10}+\ket{01}+\ket{00})$, where $\ket{11}$ is dropped out. With the same parameters that are used in the simulation shown in Fig.~\ref{fig:k_cmp}, we plot the dynamics of the new state without $\ket{11}$ in Fig.~\ref{fig:wo_11}. It is evident from this figure that the results from both master equations as expected are well matched.

Moreover, in Fig.~\ref{fig:k_cmp}, although the approximate dynamics do not mimic the exact ones in the transient time, they do match well in the steady state regime. Furthermore, even if $\ka$ is increased to a higher value, the dynamics given by both master equations still match in the final state. 

\begin{figure}[thb]
\centering
\includegraphics[width=8.6cm]{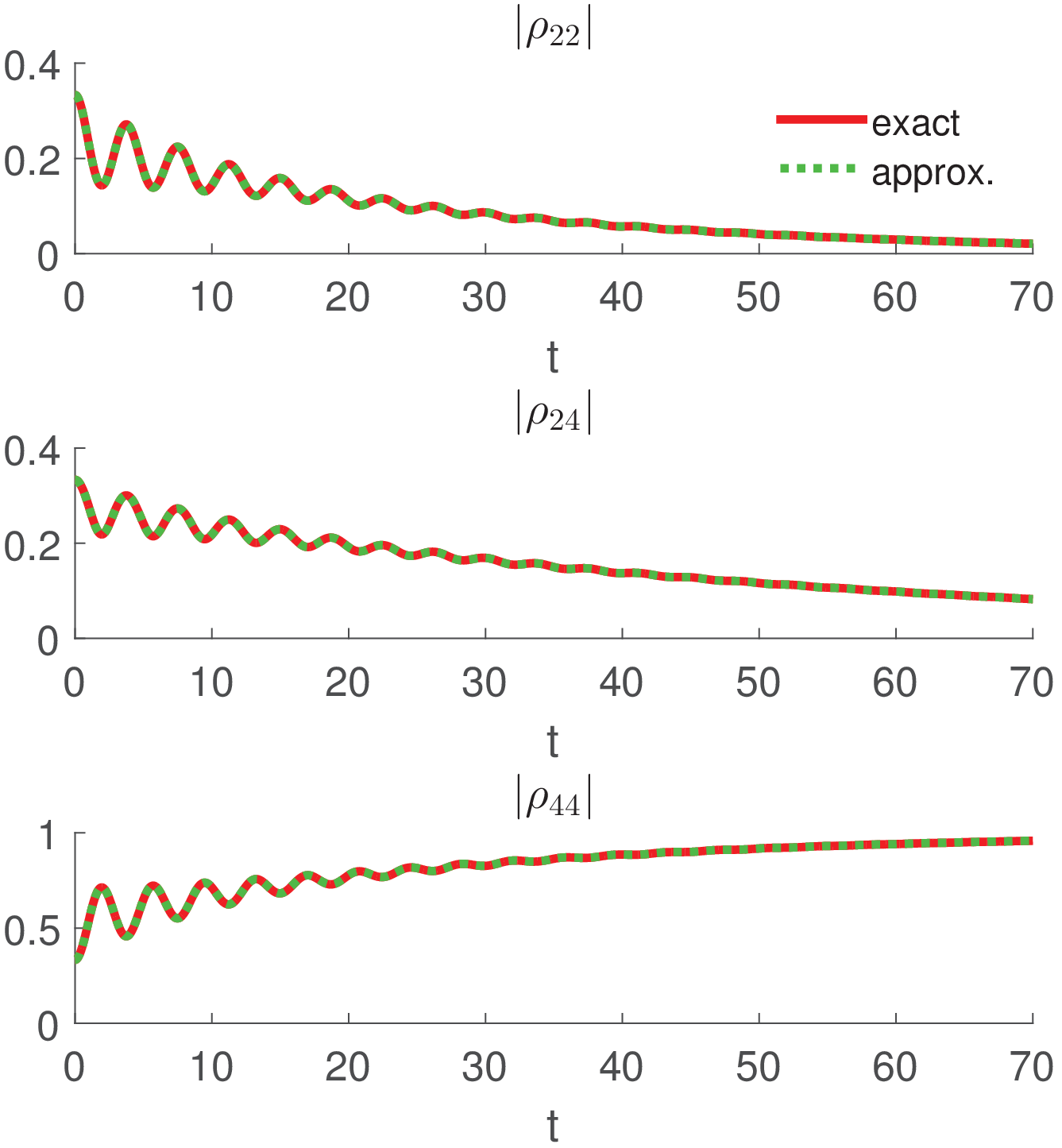}
\caption{(Color online) Dynamics of the density matrix with $\psi_0 = \f{1}{\sqrt{3}}(\ket{10}+\ket{01}+\ket{00})$. Parameters are chosen as $\gamma = 0.1$, $\w = 0.5$, $\ka = 2$, $J_{xy} = 0.7$, and $J_z = 0.3$. The other three independent elements $\rho_{11}$, $\rho_{12}$, and $\rho_{14}$ are always zero.}
\label{fig:wo_11}
\end{figure}

From the above discussions, we can summarize the findings as follows: the master equation without the terms related to the first-order noise can be quite a good approximation, especially when state $\ket{11}$ does not form part of the initial state, or at least, the exact master equation can be asymptotically approximated by neglecting the effect from the first-order noise. Hence, we can also anticipate that residual entanglement and state impurity from the initial state $\ket{10}$ would still be preserved without the first-order noise.

\section{Conclusion}
\label{sec:conclusion}
In this study, we derived the analytic form of the non-Markovian master equation of a system comprising two interacting two-level qubits coupled to a bosonic environment. Following the derivation, the master equation was validated by examining entanglement generation and state purity. The exact and approximate master equations were compared for various situations, and our results reveal that the approximate master equation is well fitted to the exact master equation in steady state, and therefore, it serves as  a good approximation in an asymptotic sense. One of the potential applications of our results is to implement two-qubit gates with this two-qubit system. Furthermore, the concept developed for this two-qubit system can be extended to multi-qubit systems.

\begin{acknowledgements}
I thank Mr. Kuan-Ting Lin for introducing me the concept of QSD, and I am also grateful to have many interesting discussions with him.

\end{acknowledgements}

\bibliography{bibl}

\begin{widetext}
\appendix
\section{{\label{app:time_deri}}Time derivatives of functions in the $O$ operator and in the exact master equation}
The time derivatives of the coefficients $f_j$ ($j = 1 \sim 5$) for the $O$ operator are as follows:
\begin{equation}
\begin{aligned}
&\f{\partial f_1}{\partial t}(t,s) = (2i\w_A + \ka_A \bar{f}_1 + \ka_B \bar{f}_3)f_1 + (-iJ_{xy} - \ka_B \bar{f}_1 + \ka_B \bar{f}_4)f_3 + (2i J_z + \ka_A \bar{f}_4 + \ka_B \bar{f}_3)f_4 -i\ka_B\bar{f}_5(t,s), \\
&\f{\partial f_2}{\partial t}(t,s) = (2i\w_B + \ka_A \bar{f}_4 + \ka_B \bar{f}_2)f_2 + (-iJ_{xy} - \ka_A \bar{f}_2 + \ka_A \bar{f}_3)f_4 + (2i J_z + \ka_A \bar{f}_4 + \ka_B \bar{f}_3)f_3 -i\ka_A\bar{f}_5(t,s), \\
&\f{\partial f_3}{\partial t}(t,s) = (2i\w_B + \ka_A \bar{f}_4 + \ka_B \bar{f}_2)f_3 + (-iJ_{xy} - \ka_A \bar{f}_2 + \ka_A \bar{f}_3)f_1 + (2i J_z + \ka_A \bar{f}_4 + \ka_B \bar{f}_3)f_2 -i\ka_A\bar{f}_5(t,s), \\
&\f{\partial f_4}{\partial t}(t,s) = (2i\w_A + \ka_A \bar{f}_1 + \ka_B \bar{f}_3)f_4 + (-iJ_{xy} - \ka_B \bar{f}_1 + \ka_B \bar{f}_4)f_2 + (2i J_z + \ka_A \bar{f}_4 + \ka_B \bar{f}_3)f_1 -i\ka_B\bar{f}_5(t,s), \\
&\f{\partial f_5}{\partial t}(t,s,s_1) = [2i(\w_A + \w_B) + \ka_A(\bar{f}_1+\bar{f}_4) + \ka_B(\bar{f}_2 + \bar{f}_3)]f_5 + (\ka_A f_1 + \ka_B f_2 - \ka_B f_3 - \ka_A f_4)\bar{f}_5(t,s_1),
\end{aligned}
\end{equation}
where $\bar{f}_j(t) = \int_0^t \mathrm{d}s~ \alpha(t,s)f_j(t,s)$ ($j = 1 \sim 4$) and $\bar{f}_5(t,s_1) = \int_0^t \mathrm{d}s~ \alpha(t,s)f_5(t,s,s_1)$. 
The initial conditions for these five functions are as follows:
\begin{equation}
f_1(t,t) = \ka_A; ~ f_2(t,t) = \ka_B; ~ f_3(t,t) = 0; ~ f_4(t,t) = 0; ~ f_5(t,t,s_1) =0; ~ f_5(t,s,t) = -i[\ka_A f_3(t,s)+\ka_B f_4(t,s)].
\end{equation}

Next, we consider the time derivatives of $\bar{f}_j(t)$ ($j = 1 \sim 4$), whose formal representations are expressed by
\begin{equation}
\label{eq:formal_fj}
\begin{aligned}
\f{\partial \bar{f}_j}{\partial t} &= \alpha(t,t) f_j(t,t) + \int_0^t \D s~\f{\partial \alpha(t,s)}{\partial t} f_j(t,s) + \int_0^t \D s~\alpha(t,s) \f{\partial f_j(t,s)}{\partial t} \\
&= \f{\gamma}{2}f_j(t,t) -\gamma \int_0^t \D s~\alpha(t,s) f_j(t,s) + \int_0^t \D s~\alpha(t,s) \f{\partial f_j(t,s)}{\partial t}.
\end{aligned}
\end{equation}
Note that the time derivatives of $\bar{f}_5(t,s_1)$ and $\tilde{f}_5(t)$ have similar forms to Eq.~(\ref{eq:formal_fj}) with $\{f_j(t,t), f_j(t,s)\}$ replaced by $\{f_5(t,t,s_1), f_5(t,s,s_1)\}$ and $\{\bar{f}_5(t,t),\bar{f}_5(t,s)\}$ respectively, where $\bar{f}_5(t,t)$ is given by
\begin{equation}
\bar{f}_5(t,t) = \int_0^t \D s~\alpha(t,s) [-i(\ka_A f_3 + \ka_B f_4)] = -i(\ka_A \bar{f}_3 + \ka_B \bar{f}_4).
\end{equation}
According to Eq.~(\ref{eq:formal_fj}), the time derivatives of $\bar{f}_j$ ($j = 1 \sim 5$) and $\tilde{f}_5$ are
\begin{equation}
\label{eq:dfbar_dt}
\begin{aligned}
&\f{\D \bar{f}_1}{\D t}(t) = \f{\gamma \ka_A}{2} + (-\gamma + 2i\w_A + \ka_A \bar{f}_1)\bar{f}_1 + (-iJ_{xy} + \ka_B \bar{f}_4)\bar{f}_3 + (2i J_z + \ka_A \bar{f}_4 + \ka_B \bar{f}_3)\bar{f}_4 -i\ka_B\tilde{f}_5(t,s),\\
&\f{\D \bar{f}_2}{\D t}(t) = \f{\gamma \ka_B}{2} + (-\gamma + 2i\w_B + \ka_B \bar{f}_2)\bar{f}_2 + (-iJ_{xy} + \ka_A \bar{f}_3)\bar{f}_4 + (2i J_z + \ka_A \bar{f}_4 + \ka_B \bar{f}_3)\bar{f}_3 -i\ka_A\tilde{f}_5(t,s),\\
&\f{\D \bar{f}_3}{\D t}(t) = (-\gamma + 2i\w_B + \ka_A \bar{f}_4 + \ka_B \bar{f}_2)\bar{f}_3 + (-iJ_{xy} - \ka_A \bar{f}_2 + \ka_A \bar{f}_3)\bar{f}_1 + (2i J_z + \ka_A \bar{f}_4 + \ka_B \bar{f}_3)\bar{f}_2 -i\ka_A\tilde{f}_5(t,s), \\
&\f{\D \bar{f}_4}{\D t}(t) = (-\gamma + 2i\w_A + \ka_A \bar{f}_1 + \ka_B \bar{f}_3)\bar{f}_4 + (-iJ_{xy} - \ka_B \bar{f}_1 + \ka_B \bar{f}_4)\bar{f}_2 + (2i J_z + \ka_A \bar{f}_4 + \ka_B \bar{f}_3)\bar{f}_1 -i\ka_B\tilde{f}_5(t,s), \\
&\f{\partial \bar{f}_5}{\partial t}(t,s_1) = [-\gamma + 2i(\w_A + \w_B) + 2\ka_A\bar{f}_1 + 2\ka_B\bar{f}_2]\bar{f}_5, \\
&\f{\D \tilde{f}_5}{\D t}(t) = -i\f{\gamma}{2}(\ka_A\bar{f}_3 + \ka_B\bar{f}_4) + [-2\gamma + 2i(\w_A + \w_B) + 2\ka_A \bar{f}_1 + 2\ka_B\bar{f}_2] \tilde{f}_5.
\end{aligned}
\end{equation}

As for the time derivatives of $F_j(t)$ ($j = 1 \sim 4$), we may also write down their formal representations as follows:
\begin{equation}
\label{eq:formal_Fj}
\begin{aligned}
\f{\D F_j}{\D t} =& \int_0^t \D s_2\alpha(t,s_2)f_j(t,s_2)\bar{f}_5^\ast(t,t) + \int_0^t \D s_1 \alpha(s_1,t)f_j(t,t)\bar{f}_5^\ast(t,s_1) \\ &+ \int_0^t \D s_1 \int_0^t \D s_2 \alpha(s_1,s_2) \f{\partial f_j}{\partial t}(t,s_2) \bar{f}_5^\ast(t,s_1) + \int_0^t \D s_1 \int_0^t \D s_2 \alpha(s_1,s_2) f_j(t,s_2) \f{\partial \bar{f}_5^\ast}{\partial t}(t,s_1), 
\end{aligned}
\end{equation}
where the first two terms equate to $\bar{f}_j(t)\bar{f}_5^\ast(t,t)$ and $f_j(t,t)\tilde{f}_5^\ast(t)$ respectively. For $\partial F_5/\partial t$, we just replace each $f_j$ term in Eq.~(\ref{eq:formal_Fj}) with $\bar{f}_5$.

Below, we list the time derivatives of $F_j(t)$ ($j = 1 \sim 5$):
\begin{equation}
\begin{aligned}
\f{\D F_1}{\D t} =&~ i\bar{f}_1(\ka_A\bar{f}_3^\ast+\ka_B\bar{f}_4^\ast) + \ka_A \tilde{f}_5^\ast + (\ka_A \bar{f}_1 + \ka_B \bar{f}_3 - \gamma -2i\w_B + 2\ka_A\bar{f}_1^\ast + 2\ka_B\bar{f}_2^\ast) F_1 + (-iJ_{xy}-\ka_B\bar{f}_1+\ka_B\bar{f}_4)F_3 \\&+ (2iJ_z+\ka_A\bar{f}_4 + \ka_B\bar{f}_3)F_4 - i\ka_B F_5, \\
\f{\D F_2}{\D t} =&~ i\bar{f}_2(\ka_A\bar{f}_3^\ast+\ka_B\bar{f}_4^\ast) + \ka_B \tilde{f}_5^\ast + (\ka_A \bar{f}_4 + \ka_B \bar{f}_2 - \gamma -2i\w_A + 2\ka_A\bar{f}_1^\ast + 2\ka_B\bar{f}_2^\ast) F_2 + (-iJ_{xy}-\ka_A\bar{f}_2+\ka_A\bar{f}_3)F_4 \\&+ (2iJ_z+\ka_A\bar{f}_4 + \ka_B\bar{f}_3)F_3 - i\ka_A F_5, \\
\f{\D F_3}{\D t} =&~ i\bar{f}_3(\ka_A\bar{f}_3^\ast+\ka_B\bar{f}_4^\ast) + (\ka_A \bar{f}_4 + \ka_B \bar{f}_2 - \gamma -2i\w_A + 2\ka_A\bar{f}_1^\ast + 2\ka_B\bar{f}_2^\ast) F_3 + (-iJ_{xy}-\ka_A\bar{f}_2+\ka_A\bar{f}_3)F_1 \\&+ (2iJ_z+\ka_A\bar{f}_4 + \ka_B\bar{f}_3)F_2 - i\ka_A F_5, \\
\f{\D F_4}{\D t} =&~ i\bar{f}_4(\ka_A\bar{f}_3^\ast+\ka_B\bar{f}_4^\ast) + (\ka_A \bar{f}_1 + \ka_B \bar{f}_3 - \gamma -2i\w_B + 2\ka_A\bar{f}_1^\ast + 2\ka_B\bar{f}_2^\ast) F_4 + (-iJ_{xy}-\ka_B\bar{f}_1+\ka_B\bar{f}_4)F_2 \\&+ (2iJ_z+\ka_A\bar{f}_4 + \ka_B\bar{f}_3)F_1 - i\ka_B F_5, \\
\f{\D F_5}{\D t} =&~ 2 \text{Re}[i\tilde{f}_5(\ka_A\bar{f}_3^\ast+\ka_B\bar{f}_4^\ast)] + [-2\gamma +4\ka_A \text{Re}(\bar{f}_1) + 4\ka_B \text{Re}(\bar{f}_2)]F_5.
\end{aligned}
\end{equation}

\end{widetext}

\end{document}